# BRIDGING THE GAP BETWEEN DATA-DRIVEN AND THEORY-DRIVEN MODELLING – LEVERAGING CAUSAL MACHINE LEARNING FOR INTEGRATIVE MODELLING OF DYNAMICAL SYSTEMS

*Completed Research Paper*

David Zapata Gonzalez, Paderborn University, Paderborn, Germany,
  david.zapata@uni-paderborn.de

Marcel Meyer, Paderborn University, Paderborn, Germany,
  marcel.meyer@uni-paderborn.de

Oliver Müller, Paderborn University, Paderborn, Germany,
  oliver.mueller@uni-paderborn.de

## Abstract

*Classical machine learning techniques often struggle with overfitting and unreliable predictions when exposed to novel conditions. Introducing causality into the modelling process offers a promising way to mitigate these challenges by enhancing predictive robustness. However, constructing an initial causal graph manually using domain knowledge is time-consuming, particularly in complex time series with numerous variables. To address this, causal discovery algorithms can provide a preliminary causal structure that domain experts can refine. This study investigates causal feature selection with domain knowledge using a data center system as an example. We use simulated time-series data to compare different causal feature selection with traditional machine-learning feature selection methods. Our results show that predictions based on causal features are more robust compared to those derived from traditional methods. These findings underscore the potential of combining causal discovery algorithms with human expertise to improve machine learning applications.*

## 1    Introduction

There exist two philosophical approaches with different goals when modelling complex socio-technical systems (Breiman, 2001; Hofman et al., 2021; Shmueli, 2010). On the one hand, one can follow an explanatory approach that values the identification of theory-driven causal mechanisms. On the other hand, one can follow a predictive modelling approach that values accurate, data-driven predictions of future outcomes.

To reconcile these two approaches, researchers have called for more integrative modelling, combining the strengths of both paradigms (Hofman et al., 2021). One concrete approach for integrative modelling is Causal Machine Learning (CML) (Schölkopf, 2022). Unlike traditional predictive machine learning, which focuses primarily on prediction accuracy, CML seeks to uncover how interventions or changes in one variable influence future outcomes, enabling robust decision-making beyond correlations. Therefore, it integrates principles from causal inference (e.g., graphical causal models, potential outcomes framework) with flexible Machine Learning (ML) algorithms, such as tree-based ensembles or neural networks.

While CML has already gained traction in medicine (Feuerriegel et al., 2024), applications of CML in Information Systems and other areas of business and economics research are just appearing (Kuzmanovic et al., 2024; Tafti & Shmueli, 2020; Von Zahn et al., 2024). In this study, we investigate the potential of CML with human collaboration for predictions in a complex dynamical system by selectin causal features to the response variable. As an exemplary application area, we focus on modelling data center operations.

Recent research in CML has introduced new methods for integrating causal insights into predictive modelling. In this study we focus on the subdomain causal discovery, especially for time series data (Runge, 2018). Causal discovery enables the construction of an initial Graphical Causal Model (GCM), which can be refined and enhanced through the integration of domain expertise. Features identified as causally linked to the target are used as input for predictive machine learning models.

In this context, we formulate the following research questions:

**RQ1:** How can we combine causal discovery of a dynamical system with domain knowledge to select relevant features for ML modelling?

**RQ2:** Is the performance of ML models trained with causal features competitive in comparison to other traditional feature selection approaches?

To answer **RQ1**, we employ causal discovery methods to identify causal relationships in time-series data from a simulated data center. The resulting causal graph is refined with experts and domain knowledge from the literature to ensure accuracy and relevance. For **RQ2**, we compare the performance of ML models trained using features selected through causal discovery, traditional feature selection methods, and all available features. These models are assessed through prediction tasks and system intervention scenarios to evaluate their performance under novel conditions.

Our findings demonstrate that combining causal discovery with human domain knowledge enhances model robustness of predictions in dynamical systems. ML models trained on causal features consistently match or outperform those using non-causal methods, especially in predicting intervention outcomes. In prediction tasks, they showed an average 6% improvement and were the best-performing models in 70% of cases. In prediction after interventions, they are on average 17% better and ranked as the top models in 80% of cases.

These results highlight the benefits of incorporating causality into ML-based modelling. By distinguishing causation from correlation, causal discovery ensures more reliable feature selection. Also, models trained on causal features require fewer inputs while maintaining comparable performance, reducing computational costs. Moreover, their superior ability to predict intervention outcomes makes them particularly valuable for "what-if" scenario analysis.

The remainder of this paper is arranged as follows: Section 2 reviews related work ML-based modelling and forecasting in the context of dynamical systems and gives a background on causal feature selection for time-series data. Section 3 describes the methods for causal discovery and ML-based predictive modelling we used in this study. Section 4 explains our experimental setup and how we evaluated the different models in two scenarios, with complete test sets and with focus on interventions. Finally, in Section 5 we discuss our empirical findings and in Section 6 we acknowledge limitations of our work, give an outlook for future work, and highlight the conclusions of our study.

## 2 Background

### 2.1 Machine Learning-based Predictive Modelling in Dynamical Systems

Machine Learning (ML) techniques are widely applied in predictive modelling of dynamical systems. Multivariate time series forecasting has a long history, and a variety of machine learning methods have been developed over time (Lim & Zohren, 2021; Masini et al., 2023) also feature selection methods are well established (Guyon & Elisseeff, 2003). They have been successfully applied across diverse fields, including natural sciences like biology, climate research, and medicine; energy sciences, such as load and generation forecasting; and economics, encompassing retail, finance, and many more (Lim et al, 2021). Other recent approaches use XAI methods to enhance feature selection (Saxena et al., 2023; Van Zyl et al., 2024). While providing valuable insights, none of the studies focus on causal feature selection and the impact of interventions on model performance. Moreover, XAI methods like SHAP do not explain causality and do not answer "what-if" questions (Lundberg et al., 2021).

## 2.2 Causal Discovery and Causal Feature Selection

Traditional (non-causal) feature selection focuses on identifying features that improve predictive accuracy for a target variable and do not attempt to model cause-effect relationships, as causal understanding is not necessary for making accurate predictions in observational settings (Kumar, 2008). In contrast, causal feature selection aims to uncover the underlying causal structure of a system (Kaddour et al., 2022; Pearl, 2014), specifically to find the Markov Blanket of the response variable, which is the smallest set of variables that retain all causal information about the target (Aliferis et al., 2010; Yu et al., 2021). This approach leads to more robust models with selected features that seek to reflect the actual causal mechanisms of the data-generating process (Kumar, 2008; Yu et al., 2020, 2021).

The foundation of causal feature selection lies in Graphical Causal Models (GCMs), which visually represent causal relationships within a system using Directed Acyclic Graphs (DAGs) (Pearl, 2009). In a DAG, variables are represented as nodes, while directed edges indicate causal influences—where an edge from variable A to variable B suggests that changes in A affect B. In the context of feature selection, causal methods prioritize variables that have a direct causal link to the target variable, ensuring that selected features not only improve predictive performance but also align with the causal structure of the data (Kumar, 2008; Pearl, 2014).

There are various characteristic patterns that help analyse causal relationships between variables in a GCM. The most basic one is a *chain*, where all arrows point in the same direction, such as, $X \rightarrow Z$ or $X \rightarrow Y \rightarrow Z$. In the latter chain, $X$ and $Z$ are conditionally independent given $Y$ ($X \perp\!\!\!\perp Z \mid Y$). If we aim to analyse the effect of $X$ on $Z$, $Y$ acts as a mediator, and controlling for it (e.g., by fixing its value) would block the effect of $X$ on $Z$. A *fork* is another important pattern that represents a common cause. For instance, in $Y \leftarrow X \rightarrow Z$, $Z$ and $Y$ are conditionally independent given $X$ ($Z \perp\!\!\!\perp Y \mid X$). In this case, X is a confounder, and it creates a noncausal association between Y and Z. Controlling for the confounder blocks this association. A *collider*, where two arrows point to the same node $X \rightarrow Z \leftarrow Y$, represents a common effect. Here, $X$ and $Y$ are conditionally independent ($X \perp\!\!\!\perp Y$) and controlling for $Z$ generates a spurious association between $X$ and $Y$ (Peters et al., 2017).

Determining a causal graph solely from domain knowledge can be challenging, but Causal discovery methods can generate an initial causal graph from data, which can then serve as a framework for various downstream causal tasks such as causal identification, effect estimation, root-cause analysis, causal feature selection (Blöbaum et al., 2022; Sharma & Kiciman, 2020). There are several families of algorithms for causal discovery (Hasan et al., 2023; Molak, 2023) with the most important ones including:

- Constraint-based methods, which rely on variable independence. The most well-known algorithm in this category is the Peter-Clark (PC) algorithm (Spirtes et al., 2001).
- Score-based methods, which use a scoring function, such as the Bayesian Information Criterion (BIC) or Minimum Description Length (MDL), to determine the best graph. A widely used algorithm in this group is Greedy Equivalence Search (Chickering, 2020).
- Gradient-based methods, which treat graph search as an optimization problem. The NOTEARS algorithm is a prominent example in this category (Zheng et al., 2018).
- Functional discovery methods, based on independent component analysis, where variables are expressed as functions of other variables and noise. An example is the NonGaussian Acyclic Model (LiNGAM) (Shimizu, 2014).

The integration of causal discovery and ML is increasingly applied in modelling physical systems. He et al. (2019) demonstrated a causality-based feature selection method that improved prediction accuracy of power distribution events by 5% and reduced computational time compared to using all features or using a traditional feature selection method. Similarly, Beucler et al., (2023) used causal feature selection to model western pacific tropical cyclones and found that they are superior to other methods to generalize over unseen cases. Also, X. Chen et al. (2022) introduced a framework for discovering causal relationships in energy-efficient building design, highlighting that causal discovery integrates domain knowledge and outperforms purely data-driven ML for answering what-if questions. These articles provide important insights but do not deal with causal feature selection with human collaboration for time-series data.

## 2.3   Causal Feature Selection and Prediction in Time Series

GCMs are typically used for cross-sectional data. Yet, time series data can also be represented with stationary DAGs (Runge et al., 2023) where the causal relationships between variables do not change over time. Figure 1 (a) shows a summary graph, also called a process graph, in which the edges are annotated with numbers indicating time lags in causal effects. The number under each node indicates that the variables also cause themselves, with lag 1. Figure 1 (b) depicts a time-series graph of the same causal links, in which time lags are columns. In this format, the acyclic nature of DAGs is more apparent.

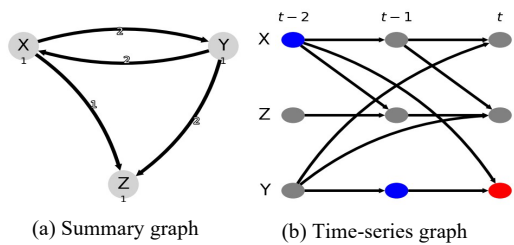

(a) Summary graph     (b) Time-series graph

*Figure 1.      Graphs to represent causality in time series.*

In this paper, we use a variant of the constrained-based PC algorithm, as it has shown good performance for causal feature selection in the context of time series problems in past studies (Assaad et al., 2022; Runge, 2020). The PC algorithm (Spirtes et al., 2001) builds upon the faithfulness assumption, stating that all observed statistical independencies in the data reflect the true underlying causal structure of the underlying data-generating process. The algorithm first establishes a fully connected undirected graph and subsequently performs iterative conditional independence tests to delete edges from the initial graph. Specifically, if two variables are conditionally independent given a set of other variables, the edge between them is removed. Finally, the algorithm orients the edges by identifying colliders (i.e., the only type of relationship where an independence test can determine the direction of the arrows) and disallowing cyclic structures (Glymour et al., 2019). The PC algorithm can be used with different types of conditional independent tests.

In time series data, every variable has an index representing a time lag. Using the PC algorithm directly on such datasets is computationally expensive (every combination of variable and time lag is considered as an individual predictor) and can lead to high false positive rates. To overcome these problems, a variant of the original PC algorithm called the $PC_1$, has been developed that performs the iterative independence test only on the most relevant lagged variables (Beucler et al., 2023; Runge et al., 2019).

Causal methods for time series rely on several key assumptions. These include the **Causal Markov Condition**, which states that, except for its direct descendants, a variable is independent of all others given its direct causes; **causal sufficiency**, which assumes there are no hidden or unobserved confounders; and **stationarity**, which requires that the statistical and causal properties of the data remain constant over time. In practice, it is often difficult to avoid violations of these assumptions, leading to potential false-positive or false-negative causal links (Assaad et al., 2022; Hasan et al., 2023; Moraffah et al., 2021; Runge, 2018).

To illustrate the difference between traditional data-driven predictive modelling and using causal feature selection from causal discovery, consider the example shown in Figure 1 (b). If the goal is to predict the variable $Y$ at time $t$ (red), a data-driven approach would ignore the causal structure of the system, feeding a combination of all observed variables and their time lags as features into an ML model. In this case, the model might identify $Z_{t-1}$ as an important feature, even though its association with $Y_t$ is spurious, caused by the confounding effect of $X_{t-2}$. In contrast, a causal prediction approach would only include the causally related variables from previous lags as predictors (blue) and disregard unrelated variables (e.g., $Z$) and time lags (e.g., $t$). This method is likely to enhance the robustness of the predictive model, particularly in scenarios involving data distribution shifts or interventions. Additionally, feeding causal variables as features into flexible, non-parametric ML models, as opposed to traditional statistical approaches (e.g., additive linear models), typically allows for more accurate models that account for non-linearities and interactions (Pearl, 2009).

## 3 Methodology

Our proposed approach for robust causal predictions involves three phases and can be seen in Figure 2: (1) causal discovery to identify variables and time lags that are possibly causally related to the response, (2) in which the resulting GCM is checked and corrected with domain knowledge and (3) using the identified variable-lag combinations as features in a traditional supervised ML model.

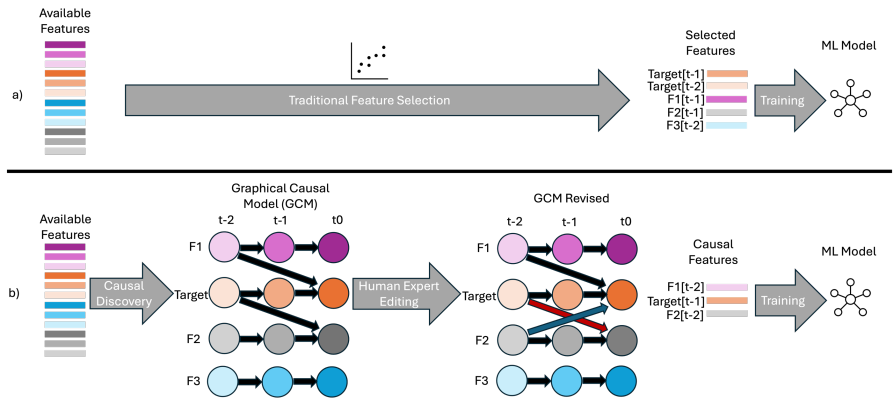

*Figure 2.    Traditional Feature Selection Framework (a) vs. Causal Feature Selection Framework (b).*

In the first phase, we use the Python library Tigramite (time-series graph-based measures of information transfer), a causal discovery framework for time-series data that relies on conditional independence tests (Runge, 2018). Following the causal inference method selector from Runge et al. (2023), we first test for stationarity using the Augmented Dickey-Fuller test from the Statsmodels library (Seabold & Perktold, 2010) with a p-value threshold of 0.01. If necessary, we apply differencing (i.e., subtracting the previous value from the current one) to achieve stationarity. Assuming a stochastic system with no hidden confounders, we employ the $PC_1$ algorithm with partial correlation tests and a significance level of 0.01 to discover the strongest causal links in the GCM (Beucler et al., 2023). For the $PC_1$ algorithm, we specify a maximum lag $\tau_{max}$ for performing the causality tests, using the average maximum lagged unconditional dependencies (lagged correlations) between all variables within a specific time period (Runge et al., 2019; Saetia et al., 2021).

In the second phase, we visualize the resulting GCM from phase (1) and refine it based on domain knowledge. The resulting corrected GCM is fed back into the pipeline. To extract causal features from the GCM, we use two approaches. In the **Causal-lags** approach, we select only those variable-lag combinations that are present in the GCM. In the **Causal-all** approach, we select all variables with at least one causal link to the target, including all lags for those variables. As a baseline, we use all variables with all lags.

Additionally, we apply various traditional feature selection methods from the scikit-learn library (Pedregosa et al., 2011) for comparison: Recursive Feature Elimination (RFE), which uses a LinearRegression to remove the least important features recursively; Principal Component Analysis (PCA), that reduces the dimensionality by transforming the features into a set of linearly uncorrelated components and features are selected based on the importance derived from the PCA which explain at least 85% of the variance; Treebased (Tree), which uses a RandomForestRegressor to estimate feature importance and we select features with importance scores above 0; Lasso with an alpha value of 0.1, which performs L1 regularization to penalize the absolute size of the coefficients leading to some coefficients being exactly zero, and we only use the variables with non-zero coefficients.

Lastly, in the third phase, we input the selected features from the training set into several regression models, including Linear Regression (LR) and Multilayer Perceptron Regressor (MLP) from the Scikit-learn library (Pedregosa et al., 2011), XGBRegressor (XGB) from the Xgboost library (T. Chen & Guestrin, 2016), and LGBMRegressor (LGBM) from the LightGBM library (Ke et al., 2017). For all models except Linear Regression, we perform hyperparameter tuning using random search and cross-validation ($K = 3$). Finally, we evaluate the predictive accuracy of the models on the test set. Figure 3 provides an overview of our pipeline for (causal) feature selection, ML training, and evaluation.

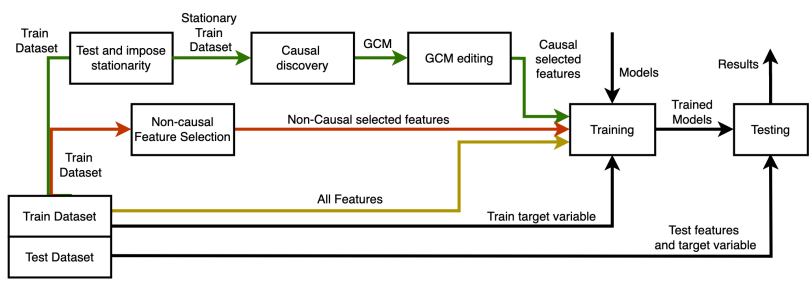

*Figure 3.    Data pipeline illustrating the information flow. Green indicates the causal feature selection path, orange represents the non-causal feature selection, and yellow for all features. Black denotes data and models used across all methods.*

## 4    Experiments

### 4.1    Simulation

Data centers are pivotal to the global digital economy, underpinning the growing adoption of AI and related technologies (Liu et al., 2020; Masanet et al., 2020). In 2016, data centers consumed about 200 Terawatt-hours of electricity - equivalent to the annual consumption of South Africa (Worldometers.info, 2024). By 2030, this demand is projected to increase two- to threefold (Koot & Wijnhoven, 2021) paralleling the energy consumption forecast for the entire African continent (Ahmad & Zhang, 2020). Optimizing data center operations, therefore, represents a critical and timely challenge while offering an ideal testbed for studying the role of causality in ML models.

We use EnergyPlus and Sinergym to simulate the operations of a data center (Jiménez-Raboso et al., 2021). EnergyPlus simulation models have been widely validated (Crawley et al., 2001), and Sinergym enhances usability and reproducibility by integrating with Python and its data science libraries (Manjavacas et al., 2024; Zhang et al., 2019).

The EnergyPlus simulation represents an air-cooled data center with no aisle containment and two asymmetrical standalone zones (East and West). For this study, we focus only on the West zone, which has a total area of 232.26 m². The zone includes a single air loop system, a Heating, Ventilation, and Air Conditioning (HVAC), an outdoor air system, variable air volume fans, direct and indirect evaporative coolers, direct expansion cooling coils, and no windows (Moriyama et al., 2018). The dominant heat source is the Information Technology Equipment (ITE), although other marginal heat sources, such as illumination, also contribute (Li et al., 2019).

We simulate two full years of data center operations. Each time step in the simulation corresponds to one hour. The simulations use New York weather data from the EnergyPlus weather files. The simulation is repeated 100 times with variations in both weather conditions and IT loads to provide 100 datasets.

Interventions and changes in a dynamical system in a real-world environment could lead to business-critical impacts and allows the possibility of investigating the what-if scenarios. Therefore, we introduced several interventions during each simulation by adjusting the temperature cooling setpoint (*Cool_Set*), on average 16 times for the two-year period, the setpoints are not the same in the first year as in the second. This is designed to test how well the different predictive models generalize to what-if scenarios and abrupt changes in the dynamical system. We use the simulate data therefore for prediction tasks and we evaluate the models in two forms: first for predictions of all the values in the test set to evaluate how the models perform in a normal prediction task (Section 4.4) and secondly for prediction after change points on the test set to evaluate how the models perform when exposed to novel conditions (Section 4.5).

## 4.2 Exploratory Data Analysis

We split the data from each simulation run into a training set (the first year, 50% of the data) and a test set (the second year, 50% of the data), ensuring a full cycle of four seasons and avoiding temporal leakage. Figure 4 (a) illustrates the time series for the training set of the first simulation run, the x-axis shows the timesteps while Figure 4 (b) lists the variables along with their units of measurement and summary statistics. Many variables exhibit non-stationary behaviour. Therefore, we tested all series for stationarity and converted them to stationary series, when necessary, as this is a prerequisite for time-series causal discovery. Some time series, such as those involving interventions in the *Cool_set,* show rapid shifts in mean.

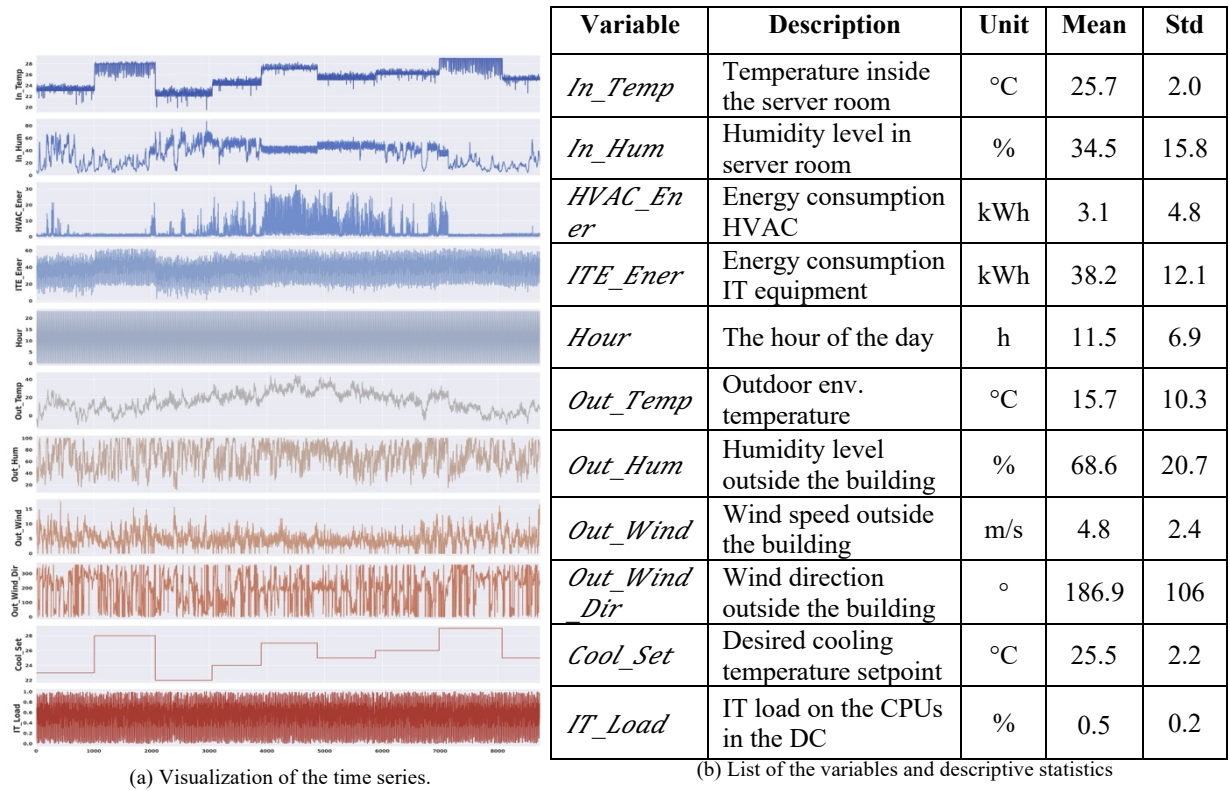

| Variable | Description | Unit | Mean | Std |
|---|---|---|---|---|
| *In_Temp* | Temperature inside the server room | °C | 25.7 | 2.0 |
| *In_Hum* | Humidity level in server room | % | 34.5 | 15.8 |
| *HVAC_Ener* | Energy consumption HVAC | kWh | 3.1 | 4.8 |
| *ITE_Ener* | Energy consumption IT equipment | kWh | 38.2 | 12.1 |
| *Hour* | The hour of the day | h | 11.5 | 6.9 |
| *Out_Temp* | Outdoor env. temperature | °C | 15.7 | 10.3 |
| *Out_Hum* | Humidity level outside the building | % | 68.6 | 20.7 |
| *Out_Wind* | Wind speed outside the building | m/s | 4.8 | 2.4 |
| *Out_Wind_Dir* | Wind direction outside the building | ° | 186.9 | 106 |
| *Cool_Set* | Desired cooling temperature setpoint | °C | 25.5 | 2.2 |
| *IT_Load* | IT load on the CPUs in the DC | % | 0.5 | 0.2 |

(a) Visualization of the time series.     (b) List of the variables and descriptive statistics

*Figure 4.  Data exploration and explanation of variables in the training set of the first simulation.*

Furthermore, other variables, particularly the *IT_Load,* display strong cyclical patterns, while the *Out_Temp* reflects clear seasonal trends. The *In_Temp*, and to a lesser degree *ITE_Ener,* fluctuate with changes in the *Cool_set.* Additionally, *IT_Load, In_Temp,* and *ITE_Ener* appear to vary according to the time of day. The *HVAC_Elec* demonstrates a marked increase in energy consumption during summer months, when higher *Out_Temp* values require the cooling system to use compressors more intensively, resulting in increased energy use.

## 4.3 Causal Discovery and domain knowledge

We perform causal discovery in the training set of the first simulation, the causal mechanisms are the same for the all the simulations. As outlined in Section 3, we then calculated the maximum lagged correlation between all variables. We tested lags from 1 to 24 and determined the average maximum correlation, $\tau_{max}$, to be 9. Using this $\tau_{max}$ and a significance level (alpha) of 0.01, we ran the PC$_1$ to generate the initial GCM.

Figure *5* (a) displays the resulting initial summary graph. The edge colors represent partial correlations between variables, while the node colors indicate autocorrelation. This initial GCM provides insights into the causal structure of the data-generating process. For example, we can observe that *Cool_set* affects *In_Temp* after two time steps. However, an initial GCM might now be without errors due to violations of the causal discovery assumptions and it is important to align it with the domain knowledge in the field. We showed the first causal graph to experts and made minor adjustments by adding or removing links. For instance, it is implausible for *Out_Hum* or *ITE_Ener* to cause the *Hour* of the day, or for *HVAC_Ener* to cause *IT_Load*. Additionally, some links are missing, we know for example, that *In_Temp* causally affects *ITE_Ener* because higher temperatures increase the energy consumption of built-in CPU fans (Sun et al., 2021) and it similarly impacts *HVAC_Ener.* Moreover, *Out_Temp* affects both *In_Temp* and *HVAC_Ener* (Silva-Llanca et al., 2023).

In a static causal graph, editing simply involves adding a link between two variables. However, in a time-series causal graph, we must also specify the appropriate lag. In our case, we select the first lag, as causal effects typically diminish over time, and a data center controlled to maintain a certain *In_Temp* will return to stability quickly after any disturbance. This can be observed in Figure 5 (a) where changes to *Cool_set* affect *In_Temp*.

Figure *5* (b) shows the final edited graph. Newly added links are marked in blue, while removed links are marked in red. Although this graph may still not be perfect, it represents the best possible encoding of the system's causal structure based on the data and business knowledge available.

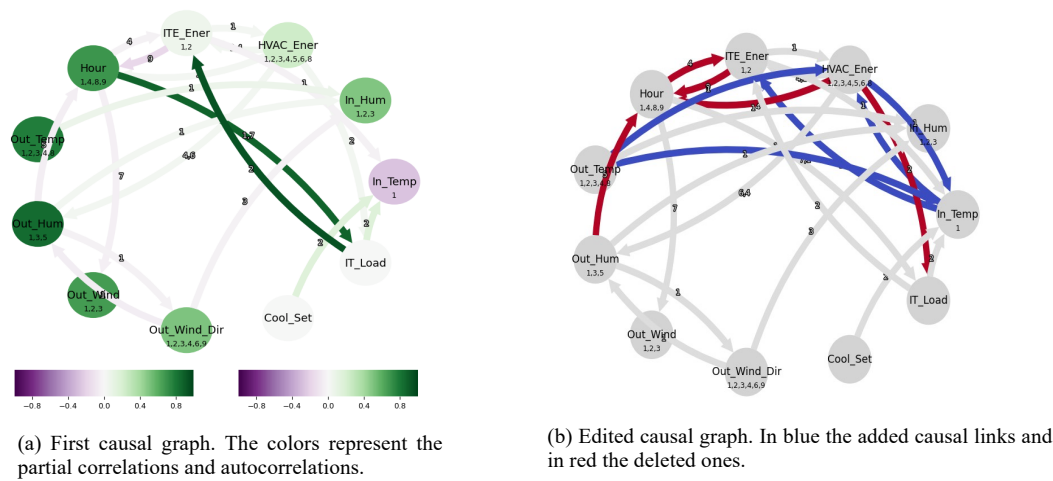

(a) First causal graph. The colors represent the partial correlations and autocorrelations.

(b) Edited causal graph. In blue the added causal links and in red the deleted ones.

*Figure 5.*     *Graphical causal models of the data center operations.*

While obtaining a perfect causal graph is typically unfeasible in practical scenarios give that assumptions of causal discovery might be violated, generating an initial graph with causal discovery and refining it with domain knowledge drastically reduces the time required for feature selection. Without a causal graph, feature selection based solely on domain knowledge would require in this case evaluating 99 features (11 Variables and 9 timesteps) . A task that would be time-consuming or even impossible in practice for experts, as they may understand which features influence others but may not know the exact lag—for example, that *cool_set* affects *In_Temp* two hours after a setpoint change. In contrast, an initial causal discovery provides a structured starting point, allowing experts to refine a much smaller subset of features and make targeted adjustments to the graph with a marginal time investment.

## 4.4 Training and Evaluating ML-based Predictive Models

We selected two different response variables, *ITE_Ener* and *In_Temp*, since they are commonly used in the literature (Lin et al., 2022; Liu et al., 2020; Tabrizchi et al., 2023) and trained various regression models using both causal and non-causal features as input. Figure 6 illustrates this process. For the models using the Causal-lag feature selection method (Figure 6 (a)), we selected only the causally related variable-lag combinations (as determined by the final edited GCM) as predictors. In contrast, we used all available variable-lag combinations for models with all predictors (Figure 6 (b)).

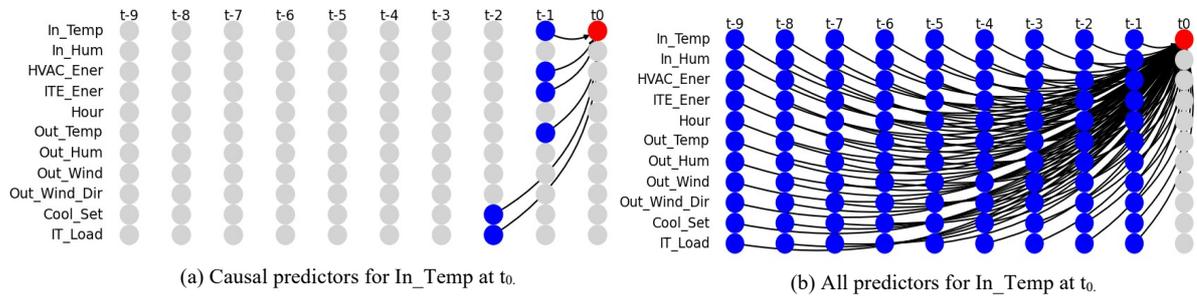

(a) Causal predictors for In_Temp at $t_0$.  (b) All predictors for In_Temp at $t_0$.

*Figure 6.  Time-series graph with predictors.*

We begin by comparing the performance of causal versus non-causal feature selection methods over the complete 1-year test set. Each model is trained and evaluated individually for each simulation run, and the results are averaged across all 100 runs. To compare the models, we use the Mean Absolute Error (MAE) as the primary metric, $MAE = \frac{1}{n}\sum_{i=1}^{n}|y_i - \hat{y}|$, where $y_i$ is the actual value of the target variable for index "i" in the test set, and $\hat{y}_i$ is the predicted value. Also, we use the Mean Absolute Percentage Error (MAPE), $MAPE = \frac{100}{n}\sum_{i=1}^{n}\left|\frac{y_i-\hat{y}_i}{y_i}\right|$, which provides a percentage-based measure of predictive accuracy. Our baseline model is a Linear Regression (LR) model that uses all available features.

Table 1 presents the results of our experiments for the *ITE_Ener* response variable, sorted by MAE. The Causal-all feature selection method combined with an MLP clearly outperformed all other models across all metrics, using 36 out of the 99 available features. Figure 7 (a) shows the frequency with which each feature selection method produced the best model for *ITE_Ener* in the 100 simulations: Causal-all was the best 55 times, followed by Causal-lags with 18 and RFE with 15.

Table 2 presents the results for the *In_Temp* response variable. In this case, the two causal feature selection methods -Causal-lags with only 6 features and Causal-all with 54 features- occupy the top three spots. Interestingly, the underlying ML models used for the three best-performing solutions are different: MLP, LR, and XGB. Figure 7 (b) shows that the Causal-lags feature selection method performed the best in 58 simulations, followed by Causal-all with 21 and RFE with 11.

In summary, at this point, it can be concluded that across both prediction tasks, feature selection methods based on causal discovery (i.e., Causal-lags or Causal-all) outperformed other approaches in more than 70% of the simulation runs.

| Feat. Select. | Mod | Nº F | MAE | MAPE |
|---|---|---|---|---|
| Causal-all | MLP | 36 | 2.223 | 6.998 |
| RFE | MLP | 49 | 2.520 | 7.872 |
| Causal-lags | MLP | 6 | 2.616 | 8.964 |
| Lasso | MLP | 81 | 2.626 | 8.146 |
| All | MLP | 99 | 2.637 | 8.173 |
| Causal-all | LGBM | 36 | 2.654 | 8.420 |
| Lasso | XGB | 81 | 2.689 | 8.510 |
| RFE | XGB | 49 | 2.690 | 8.520 |
| Lasso | LGBM | 81 | 2.700 | 8.558 |
| All | XGB | 99 | 2.705 | 8.588 |
| Causal-all | XGB | 36 | 2.708 | 8.614 |
| All | LGBM | 99 | 2.747 | 8.753 |
| RFE | LGBM | 49 | 2.766 | 8.800 |
| Tree | MLP | 6 | 2.812 | 8.984 |
| Causal-lags | LGBM | 6 | 2.820 | 9.663 |
| Tree | LGBM | 6 | 2.873 | 9.229 |
| Tree | XGB | 6 | 2.874 | 9.251 |
| Causal-lags | XGB | 6 | 2.916 | 10.051 |
| Causal-lags | LR | 6 | 3.170 | 11.060 |
| Causal-all | LR | 36 | 3.349 | 10.846 |
| RFE | LR | 49 | 3.358 | 10.866 |
| All | LR | 99 | 3.362 | 10.871 |
| Lasso | LR | 81 | 3.365 | 10.879 |
| Tree | LR | 6 | 3.435 | 11.230 |
| PCA | LGBM | 6 | 9.612 | 34.150 |
| PCA | XGB | 6 | 9.654 | 34.311 |
| PCA | MLP | 6 | 9.699 | 34.365 |
| PCA | LR | 6 | 10.096 | 36.010 |

*Table 1.    Prediction results for ITE_Ener.*

| Feat. Select. | Mod | Nº F | MAE | MAPE |
|---|---|---|---|---|
| Causal-lags | MLP | 6 | 0.210 | 0.831 |
| Causal-lags | LR | 6 | 0.217 | 0.863 |
| Causal-all | XGB | 54 | 0.223 | 0.887 |
| RFE | LGBM | 49 | 0.223 | 0.889 |
| Causal-all | LR | 54 | 0.224 | 0.885 |
| Causal-lags | LGBM | 6 | 0.224 | 0.892 |
| RFE | LR | 49 | 0.224 | 0.888 |
| Causal-lags | XGB | 6 | 0.224 | 0.892 |
| RFE | XGB | 49 | 0.224 | 0.893 |
| All | LR | 99 | 0.225 | 0.889 |
| Lasso | LR | 45 | 0.225 | 0.893 |
| Causal-all | LGBM | 54 | 0.228 | 0.905 |
| All | XGB | 99 | 0.229 | 0.912 |
| All | LGBM | 99 | 0.232 | 0.923 |
| Lasso | XGB | 45 | 0.234 | 0.930 |
| Lasso | LGBM | 45 | 0.239 | 0.947 |
| Tree | LR | 6 | 0.240 | 0.955 |
| RFE | MLP | 49 | 0.241 | 0.956 |
| Tree | MLP | 6 | 0.245 | 0.972 |
| Lasso | MLP | 45 | 0.249 | 0.985 |
| Causal-all | MLP | 54 | 0.251 | 0.993 |
| Tree | XGB | 6 | 0.2773 | 1.106 |
| All | MLP | 99 | 0.2862 | 1.130 |
| Tree | LGBM | 6 | 0.2866 | 1.142 |
| PCA | LGBM | 6 | 1.887 | 7.510 |
| PCA | XGB | 6 | 1.888 | 7.513 |
| PCA | MLP | 6 | 1.891 | 7.525 |
| PCA | LR | 6 | 1.898 | 7.554 |

*Table 2.    Prediction results for In_Temp.*

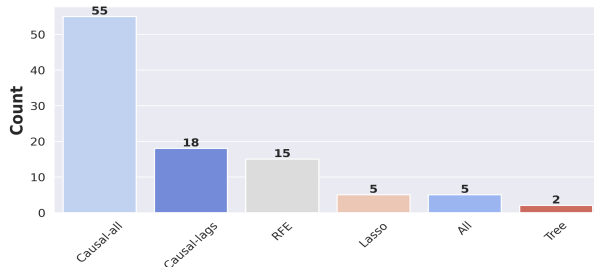
(a) Best feature selection method for ITE_Ener.

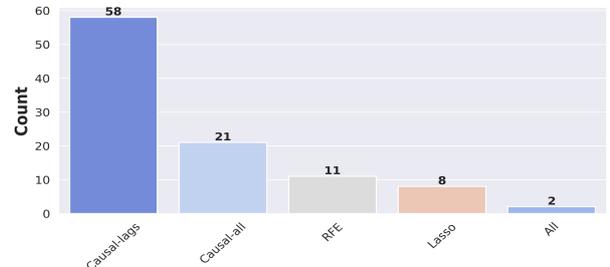
(b) Best feature selection method for In_Temp.

*Figure 7.    Number of times a feature selection had the best prediction in the simulations' data.*

## 4.5    Predictive Accuracy After Interventions

Causal inference methods are particularly effective when addressing "what-if" or counterfactual questions (Pearl, 2009). Therefore, we now shift our focus to predictions in scenarios following interventions, such as changes to the cooling set point (*Cool_set*).

In the simulation, we adjusted the *Cool_set* to observe its effect on *In_Temp*. The *Cool_set* was altered on an average of 8 times in both first year (training set) and the second year (test set). Figure 8 (a) shows the time series for *Cool_set* and *In_Temp* in the test set of the first simulation. The red markers highlight 5 points following a change in *Cool_set*. The magenta square zooms in on one of these points, as shown in Figure 8 (b). After each change in *Cool_set*, the *In_Temp* begins to change after approximately two-timesteps. Subsequently, *In_Temp* experiences some fluctuations before stabilizing again. While the

*Integrative Modelling with Causal Discovery*

stabilization behaviour varies between different changes in *Cool_set* and their corresponding effects on *In_Temp,* it typically occurs around five time steps after the initial change in *Cool_set*.

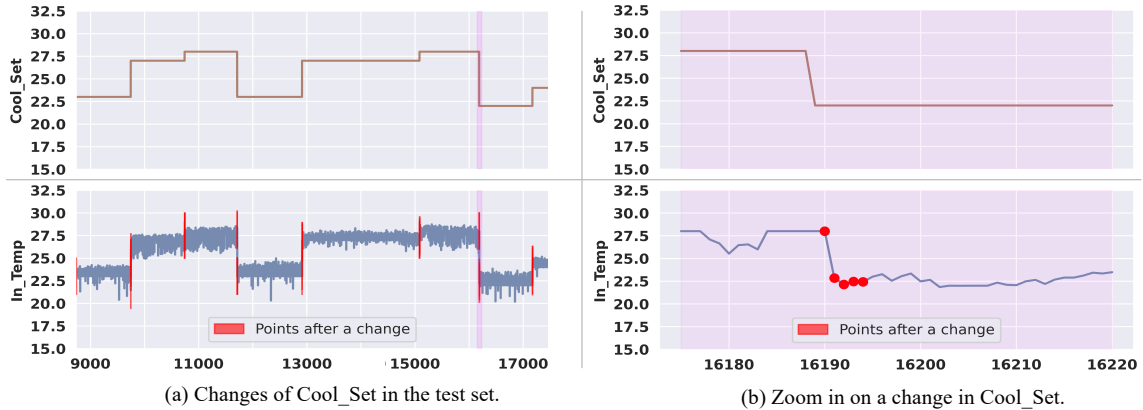

(a) Changes of Cool_Set in the test set.    (b) Zoom in on a change in Cool_Set.

*Figure 8.    Values of In_Temp after a change in Cool_set.*

To evaluate model performance after these interventions, we used the trained models from the previous task, but calculated MAE specifically for the window after each setpoint change. This is represented as $MAE_w = MAE_{t+1,...,t+5}$. We did the same for MAPE. *Table 3* shows the results and that all models performed worse after the *Cool_set* intervention, with MAEs more than doubling compared to Table 2. However, models with causal feature selection outperformed others even more clearly. Previously, they were only 6% better for the *In_Temp* target, but now they outperform other methods by about 17%.

| Feat. Select. | Model | Nº F | MAEw | MAPEw |
|---|---|---|---|---|
| Causal-lags | MLP | 6 | 0.429 | 1.704 |
| Causal-lags | XGB | 6 | 0.430 | 1.714 |
| Causal-lags | LGBM | 6 | 0.431 | 1.721 |
| Causal-lags | LR | 6 | 0.446 | 1.780 |
| RFE | LR | 49 | 0.515 | 2.056 |
| Causal-all | LR | 54 | 0.515 | 2.057 |
| All | LR | 99 | 0.516 | 2.059 |
| Lasso | XGB | 45 | 0.530 | 2.115 |
| Lasso | LR | 45 | 0.530 | 2.118 |
| Causal-all | XGB | 54 | 0.533 | 2.124 |
| Lasso | LGBM | 45 | 0.536 | 2.140 |
| Tree | LR | 6 | 0.536 | 2.146 |
| RFE | XGB | 49 | 0.537 | 2.141 |
| RFE | LGBM | 49 | 0.537 | 2.144 |
| Causal-all | LGBM | 54 | 0.539 | 2.149 |
| All | XGB | 99 | 0.539 | 2.153 |
| All | LGBM | 99 | 0.544 | 2.171 |
| Tree | LGBM | 6 | 0.560 | 2.242 |
| Tree | XGB | 6 | 0.562 | 2.248 |
| Lasso | MLP | 45 | 0.604 | 2.412 |
| Causal-all | MLP | 54 | 0.645 | 2.579 |
| Tree | MLP | 6 | 0.658 | 2.594 |
| RFE | MLP | 49 | 0.669 | 2.673 |
| All | MLP | 99 | 0.684 | 2.726 |
| PCA | XGB | 6 | 1.711 | 6.887 |
| PCA | LGBM | 6 | 1.713 | 6.890 |
| PCA | LR | 6 | 1.717 | 6.907 |
| PCA | MLP | 6 | 1.723 | 6.928 |

*Table 3.    Results for In_Temp after interventions.*

Figure 9 (a) reinforces this finding, showing that the Causal-lag feature selection method produced the best model in 80 out of the 100 simulation runs, followed by Lasso with 9. Figure 9 (b) illustrates how the average MAE for each feature selection method evolves over time following a change in the setpoint. Immediately after the change, the causal models perform significantly better than the others. However, this difference diminishes as the system stabilizes.

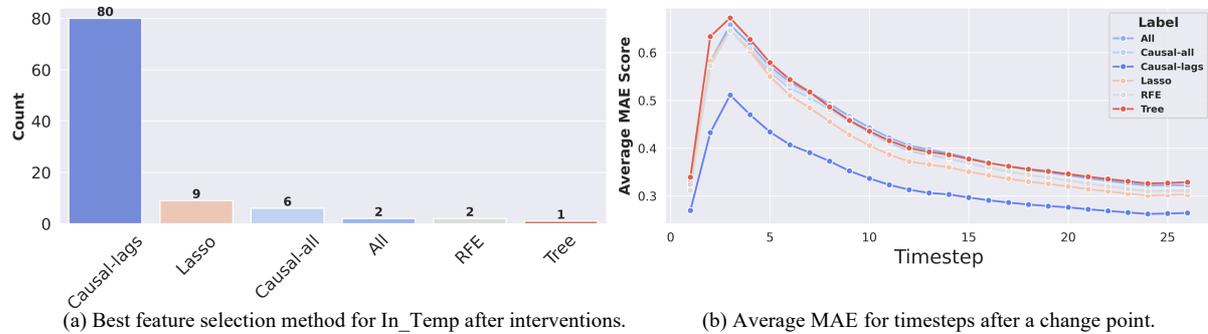

(a) Best feature selection method for In_Temp after interventions.   (b) Average MAE for timesteps after a change point.

*Figure 9.    Number of times a feature selection had the best prediction in the simulations' data for In_Temp and Average MAE after interventions*

## 5     Discussion

In this paper, we explored the use of causal discovery for understanding the causal dynamics of data center operations and selecting causally relevant features to predict energy consumption and server room temperatures. Addressing **RQ1**, we employed causal discovery techniques for time-series data to identify causal features, which were subsequently refined with domain knowledge to ensure accuracy and relevance. This hybrid approach allowed us to include critical variables that the automated process may have missed. For **RQ2**, we compared models trained on causally selected features to those trained using traditional feature selection methods and those utilizing all available features. The causal models demonstrated competitive performance in prediction tasks while excelling in scenarios involving system interventions, despite using significantly fewer features.

Our causal discovery framework integrates domain expertise early in the modelling pipeline, which is particularly valuable given the unique challenges of time-series data in contrast to static use cases (X. Chen et al., 2022). In our study, we selected, deleted links with lags based on expert knowledge, aligning the causal graph with the system's operational characteristics of data centers. Overall, incorporating causal discovery as an initial step in modelling complex dynamical systems reduced computational demands and ensured that only essential variables were used in training.

Remarkably, causal discovery-based approaches performed as well as or better than models using all available features or traditional feature selection methods. This aligns with the findings of Beucler et al. (2023) and highlights the efficiency of causal models, which require fewer inputs while maintaining strong predictive performance. By simplifying the modelling process, causal methods also reduce the environmental impact of model training, an important consideration given the computational intensity of modern ML workflows.

When predicting the effects of system interventions, such as adjusting the cooling set point (*Cool_set*) and its impact on internal temperature (*In_Temp*), causal prediction approaches consistently outperformed other methods. This robustness underscores the suitability of causal models for addressing "what-if" scenarios, counterfactuals, and other causal inquiries from the decision makers, tasks that are impractical or unreliable with traditional feature selected models (Lundberg et al., 2021). These strengths make causal modelling a powerful tool for guiding policy decisions and operational planning in complex systems.

Another contribution of this study is the exploration of the Causal-all approach, which includes all variables with at least one causal link in the GCM, along with their lags, as features. While this approach performed well for predicting IT equipment energy consumption *ITE_Ener* and internal temperature

*In_Temp*, it was less effective for intervention scenarios compared to models using only causal the Causal-lags approach. These findings emphasize the importance of targeted causal feature selection for interventional tasks, highlighting the need for fine-grained causal analysis in such contexts.

While causal discovery is highly valuable for defining causal links in a time-series graph, it is not without limitations and may contain imperfections in practical applications (Assaad et al., 2022; Hasan et al., 2023; Moraffah et al., 2021; Runge, 2018). The availability and integration of domain knowledge are essential for refining the causal discovery process. However, defining a causal graph for a complex dynamical system—particularly determining the relevant lags—can be difficult, time-consuming, or impossible, even for experts. Therefore, using causal discovery as an initial step is crucial, as it provides a foundational graph that experts can further refine, making the effort to obtain the causal graph drastically smaller. This iterative approach ensures that the final set of features is grounded in operational reality. By embedding causality within the machine learning pipeline, we enhance the robustness of predictive models, particularly for intervention scenarios.

# 6    Conclusions

This study highlights the value of incorporating causal discovery into predictive modelling for dynamic systems, such as data center operations. By identifying and refining causally relevant features with domain knowledge, we reduced the number of features needed, ensuring robust predictions even in scenarios involving system interventions.

Models trained with causal features demonstrated comparable predictive performance to traditional methods while excelling in intervention-based tasks, offering reliable insights for "what-if" analyses and counterfactual reasoning. These models simplify the modelling process, enhance transparency, and align predictions with operational realities, making them invaluable tools for informed decision-making in complex dynamical environments.

However, there are some limitations of this study which can be eliminated with further research. In our investigated use case, basic physics knowledge, literature research, and consultations with data center expert were sufficient to edit the graph. However, with many more variables, the causal discovery algorithm might be very computational expensive and the editing workload for experts could become very time-consuming steps. Furthermore, determining the appropriate lags for causal relationship in time-series data might be even with domain knowledge a challenging task as it may not always be possible to precisely determine the duration from an intervention to its effect. Another limitation is the reliance on simulation data. While using real data would be preferable, implementing and monitoring interventions in real-world scenarios is challenging, as the economic operation of the system must be preserved. Nevertheless, our results should be further validated in other complex dynamic systems and real-world settings. Moreover, it would be interesting to replicate our methodology with varying amounts of data available. We hypothesize that causal-feature based models perform better than purely data-driven methods in situations with only little data. Also, a further approach in using causality-based models can be the replacement of a simulation to train RL agents. These efforts would further validate and expand the applicability of causal methods in other dynamic and intervention-driven systems.

# 7    Acknowledgements

This study was supported by the Federal Ministry for the Environment, Nature Conservation, Nuclear Safety and Consumer Protection of Germany under Grant No. 67KI32008B (DC2HEAT - Data center HEat Recovery with AI-Technologies), and we gratefully acknowledge their support.

## References

Ahmad, T., & Zhang, D. (2020). A critical review of comparative global historical energy consumption and future demand: The story told so far. *Energy Reports*, *6*, 1973–1991.

Aliferis, C. F., Statnikov, A., Tsamardinos, I., Mani, S., & Koutsoukos, X. D. (2010). *Local Causal and Markov Blanket Induction for Causal Discovery and Feature Selection for Classification Part I: Algorithms and Empirical Evaluation*.

Assaad, C. K., Devijver, E., & Gaussier, E. (2022). Survey and Evaluation of Causal Discovery Methods for Time Series. *Journal of Artificial Intelligence Research*, *73*, 767–819. https://doi.org/10.1613/jair.1.13428

Beucler, T., Tam, F. I.-H., Gomez, M. S., Runge, J., Gerhardus, A., & others. (2023). Selecting robust features for machine-learning applications using multidata causal discovery. *Environmental Data Science*, *2*, e27.

Blöbaum, P., Götz, P., Budhathoki, K., Mastakouri, A. A., & Janzing, D. (2022). DoWhy-GCM: An extension of DoWhy for causal inference in graphical causal models. *arXiv Preprint arXiv:2206.06821*.

Breiman, L. (2001). Statistical Modeling: The Two Cultures. *Statistical Science*, *16*(3), 199–215.

Chen, T., & Guestrin, C. (2016). XGBoost: A Scalable Tree Boosting System. *Proceedings of the 22nd ACM SIGKDD International Conference on Knowledge Discovery and Data Mining*, 785–794. https://doi.org/10.1145/2939672.2939785

Chen, X., Abualdenien, J., Singh, M. M., Borrmann, A., & Geyer, P. (2022). Introducing causal inference in the energy-efficient building design process. *Energy and Buildings*, *277*, 112583. https://doi.org/10.1016/j.enbuild.2022.112583

Chickering, M. (2020). Statistically efficient greedy equivalence search. *Conference on Uncertainty in Artificial Intelligence*, 241–249.

Crawley, D. B., Lawrie, L. K., Winkelmann, F. C., Buhl, W. F., Huang, Y. J., Pedersen, C. O., Strand, R. K., Liesen, R. J., Fisher, D. E., Witte, M. J., & others. (2001). EnergyPlus: Creating a new-generation building energy simulation program. *Energy and Buildings*, *33*(4), 319–331.

Feuerriegel, S., Frauen, D., Melnychuk, V., Schweisthal, J., Hess, K., Curth, A., Bauer, S., Kilbertus, N., Kohane, I. S., & Schaar, M. van der. (2024). Causal machine learning for predicting treatment outcomes. *Nature Medicine*, *30*(4), 958–968. https://doi.org/10.1038/s41591-024-02902-1

Glymour, C., Zhang, K., & Spirtes, P. (2019). Review of Causal Discovery Methods Based on Graphical Models. *Frontiers in Genetics*, *10*, 524. https://doi.org/10.3389/fgene.2019.00524

Guyon, I., & Elisseeff, A. (2003). *An Introduction to Variable and Feature Selection*.

Hasan, U., Hossain, E., & Gani, M. O. (2023). A survey on causal discovery methods for iid and time series data. *arXiv Preprint arXiv:2303.15027*.

He, M., Gu, W., Zhou, Y., Kong, Y., & Zhang, L. (2019). Causal feature selection for physical sensing data: A case study on power events prediction. *Adjunct Proceedings of the 2019 ACM International Joint Conference on Pervasive and Ubiquitous Computing and Proceedings of the 2019 ACM International Symposium on Wearable Computers*, 565–570.

Hofman, J. M., Watts, D. J., Athey, S., Garip, F., Griffiths, T. L., Kleinberg, J., Margetts, H., Mullainathan, S., Salganik, M. J., Vazire, S., Vespignani, A., & Yarkoni, T. (2021). Integrating explanation and prediction in computational social science. *Nature*, *595*(7866), 181–188. https://doi.org/10.1038/s41586-021-03659-0

Jiménez-Raboso, J., Campoy-Nieves, A., Manjavacas-Lucas, A., Gómez-Romero, J., & Molina-Solana, M. (2021). Sinergym: A building simulation and control framework for training reinforcement learning agents. *Proceedings of the 8th ACM International Conference on Systems for Energy-Efficient Buildings, Cities, and Transportation*, 319–323. https://doi.org/10.1145/3486611.3488729

Kaddour, J., Lynch, A., Liu, Q., Kusner, M. J., & Silva, R. (2022). *Causal Machine Learning: A Survey and Open Problems* (arXiv:2206.15475). arXiv. http://arxiv.org/abs/2206.15475

Ke, G., Meng, Q., Finley, T., Wang, T., Chen, W., Ma, W., Ye, Q., & Liu, T.-Y. (2017). Lightgbm: A highly efficient gradient boosting decision tree. *Advances in Neural Information Processing Systems*, *30*.

Koot, M., & Wijnhoven, F. (2021). Usage impact on data center electricity needs: A system dynamic forecasting model. *Applied Energy*, *291*, 116798.

Kumar, V. (2008). *Chapman & Hall/CRC Data Mining and Knowledge Discovery Series*.

Kuzmanovic, M., Frauen, D., Hatt, T., & Feuerriegel, S. (2024). Causal Machine Learning for Cost-Effective Allocation of Development Aid. *Proceedings of the 30th ACM SIGKDD Conference on Knowledge Discovery and Data Mining*, 5283–5294. https://doi.org/10.1145/3637528.3671551

Li, Y., Wen, Y., Tao, D., & Guan, K. (2019). Transforming cooling optimization for green data center via deep reinforcement learning. *IEEE Transactions on Cybernetics*, *50*(5), 2002–2013.

Lim, B., & Zohren, S. (2021). Time-series forecasting with deep learning: A survey. *Philosophical Transactions of the Royal Society A: Mathematical, Physical and Engineering Sciences*, *379*(2194), 20200209. https://doi.org/10.1098/rsta.2020.0209

Lin, J., Lin, W., Lin, W., Wang, J., & Jiang, H. (2022). Thermal prediction for Air-cooled data center using data Driven-based model. *Applied Thermal Engineering*, *217*, 119207. https://doi.org/10.1016/j.applthermaleng.2022.119207

Liu, Y., Wei, X., Xiao, J., Liu, Z., Xu, Y., & Tian, Y. (2020). Energy consumption and emission mitigation prediction based on data center traffic and PUE for global data centers. *Global Energy Interconnection*, *3*(3), 272–282.

Lundberg, S., Dillon, E., LaRiviere, J., Roth, J., & Syrgkanis, V. (2021). *Be careful when interpreting predictive models in search of causal insights; SHAP latest documentation—Shap.readthedocs.io*. https://shap.readthedocs.io/en/latest/example_notebooks/overviews/Be%20careful%20when%20interpreting%20predictive%20models%20in%20search%20of%20causal%20insights.html

Manjavacas, A., Campoy-Nieves, A., Jiménez-Raboso, J., Molina-Solana, M., & Gómez-Romero, J. (2024). An experimental evaluation of deep reinforcement learning algorithms for HVAC control. *Artificial Intelligence Review*, *57*(7), 173.

Masanet, E., Shehabi, A., Lei, N., Smith, S., & Koomey, J. (2020). Recalibrating global data center energy-use estimates. *Science*, *367*(6481), 984–986.

Masini, R. P., Medeiros, M. C., & Mendes, E. F. (2023). Machine learning advances for time series forecasting. *Journal of Economic Surveys*, *37*(1), 76–111. https://doi.org/10.1111/joes.12429

Molak, A. (2023). *Causal Inference and Discovery in Python: Unlock the secrets of modern causal machine learning with DoWhy, EconML, PyTorch and more*. Packt Publishing Ltd.

Moraffah, R., Sheth, P., Karami, M., Bhattacharya, A., Wang, Q., Tahir, A., Raglin, A., & Liu, H. (2021). Causal inference for time series analysis: Problems, methods and evaluation. *Knowledge and Information Systems*, *63*(12), 3041–3085. https://doi.org/10.1007/s10115-021-01621-0

Moriyama, T., De Magistris, G., Tatsubori, M., Pham, T.-H., Munawar, A., & Tachibana, R. (2018). Reinforcement Learning Testbed for Power-Consumption Optimization. In L. Li, K. Hasegawa, & S. Tanaka (Eds.), *Methods and Applications for Modeling and Simulation of Complex Systems* (Vol. 946, pp. 45–59). Springer Singapore. https://doi.org/10.1007/978-981-13-2853-4_4

Pearl, J. (2009). *Causality*. Cambridge university press.

Pearl, J. (2014). *Probabilistic Reasoning in Intelligent Systems: Networks of Plausible Inference* (1. Aufl). Elsevier Reference Monographs.

Pedregosa, F., Varoquaux, G., Gramfort, A., Michel, V., Thirion, B., Grisel, O., Blondel, M., Prettenhofer, P., Weiss, R., Dubourg, V., Vanderplas, J., Passos, A., Cournapeau, D., Brucher, M., Perrot, M., & Duchesnay, E. (2011). Scikit-learn: Machine Learning in Python. *Journal of Machine Learning Research*, *12*, 2825–2830.

Peters, J., Janzing, D., & Schölkopf, B. (2017). *Elements of causal inference: Foundations and learning algorithms*. The MIT Press.

Runge, J. (2018). Causal network reconstruction from time series: From theoretical assumptions to practical estimation. *Chaos: An Interdisciplinary Journal of Nonlinear Science*, *28*(7), 075310. https://doi.org/10.1063/1.5025050

Runge, J. (2020). Discovering contemporaneous and lagged causal relations in autocorrelated nonlinear time series datasets. *Conference on Uncertainty in Artificial Intelligence*, 1388–1397.

Runge, J., Gerhardus, A., Varando, G., Eyring, V., & Camps-Valls, G. (2023). Causal inference for time series. *Nature Reviews Earth & Environment*, *4*(7), 487–505. https://doi.org/10.1038/s43017-023-00431-y

Runge, J., Nowack, P., Kretschmer, M., Flaxman, S., & Sejdinovic, D. (2019). Detecting causal associations in large nonlinear time series datasets. *Science Advances*, *5*(11), eaau4996. https://doi.org/10.1126/sciadv.aau4996

Saetia, S., Yoshimura, N., & Koike, Y. (2021). Constructing Brain Connectivity Model Using Causal Network Reconstruction Approach. *Frontiers in Neuroinformatics*, *15*, 619557. https://doi.org/10.3389/fninf.2021.619557

Saxena, D., Kumar, J., Singh, A. K., & Schmid, S. (2023). Performance Analysis of Machine Learning Centered Workload Prediction Models for Cloud. *IEEE Transactions on Parallel and Distributed Systems*, *34*(4), 1313–1330. https://doi.org/10.1109/TPDS.2023.3240567

Schölkopf, B. (2022). *Causality for Machine Learning* (pp. 765–804).

Seabold, S., & Perktold, J. (2010). statsmodels: Econometric and statistical modeling with python. *9th Python in Science Conference*.

Sharma, A., & Kiciman, E. (2020). DoWhy: An End-to-End Library for Causal Inference. *arXiv Preprint arXiv:2011.04216*.

Shimizu, S. (2014). LiNGAM: Non-Gaussian methods for estimating causal structures. *Behaviormetrika*, *41*(1), 65–98.

Shmueli, G. (2010). To Explain or to Predict? *Statistical Science*, *25*(3).

Silva-Llanca, L., Ponce, C., Bermúdez, E., Martínez, D., Díaz, A. J., & Aguirre, F. (2023). Improving energy and water consumption of a data center via air free-cooling economization: The effect weather on its performance. *Energy Conversion and Management*, *292*, 117344. https://doi.org/10.1016/j.enconman.2023.117344

Spirtes, P., Glymour, C., & Scheines, R. (2001). *Causation, prediction, and search*. MIT press.

Sun, K., Luo, N., Luo, X., & Hong, T. (2021). Prototype energy models for data centers. *Energy and Buildings*, *231*, 110603. https://doi.org/10.1016/j.enbuild.2020.110603

Tabrizchi, H., Razmara, J., & Mosavi, A. (2023). Thermal prediction for energy management of clouds using a hybrid model based on CNN and stacking multi-layer bi-directional LSTM. *Energy Reports*, *9*, 2253–2268. https://doi.org/10.1016/j.egyr.2023.01.032

Tafti, A., & Shmueli, G. (2020). Beyond Overall Treatment Effects: Leveraging Covariates in Randomized Experiments Guided by Causal Structure. *Information Systems Research*, *31*(4), 1183–1199. https://doi.org/10.1287/isre.2020.0938

Van Zyl, C., Ye, X., & Naidoo, R. (2024). Harnessing eXplainable artificial intelligence for feature selection in time series energy forecasting: A comparative analysis of Grad-CAM and SHAP. *Applied Energy*, *353*, 122079. https://doi.org/10.1016/j.apenergy.2023.122079

Von Zahn, M., Bauer, K., Mihale-Wilson, C., Jagow, J., Speicher, M., & Hinz, O. (2024). Smart Green Nudging: Reducing Product Returns Through Digital Footprints and Causal Machine Learning. *Marketing Science*, mksc.2022.0393. https://doi.org/10.1287/mksc.2022.0393

Worldometers.info. (2024). *South Africa Electricity*. https://www.worldometers.info/electricity/south-africa-electricity/

Yu, K., Guo, X., Liu, L., Li, J., Wang, H., Ling, Z., & Wu, X. (2021). Causality-based Feature Selection: Methods and Evaluations. *ACM Computing Surveys*, *53*(5), 1–36. https://doi.org/10.1145/3409382

Yu, K., Liu, L., Li, J., Ding, W., & Le, T. D. (2020). Multi-Source Causal Feature Selection. *IEEE Transactions on Pattern Analysis and Machine Intelligence*, *42*(9), 2240–2256. IEEE Transactions on Pattern Analysis and Machine Intelligence. https://doi.org/10.1109/TPAMI.2019.2908373

Zhang, C., Kuppannagari, S. R., Kannan, R., & Prasanna, V. K. (2019). Building HVAC Scheduling Using Reinforcement Learning via Neural Network Based Model Approximation. *Proceedings of the 6th ACM International Conference on Systems for Energy-Efficient Buildings, Cities, and Transportation*, 287–296. https://doi.org/10.1145/3360322.3360861

Zheng, X., Aragam, B., Ravikumar, P. K., & Xing, E. P. (2018). Dags with no tears: Continuous optimization for structure learning. *Advances in Neural Information Processing Systems*, *31*.